\documentclass[aps,prl,twocolumn,superscriptaddress]{revtex4}
\usepackage{color}
\usepackage{amssymb}
\usepackage{epsfig}
\usepackage{graphicx,amsmath}
\begin{document}
\title{Comment on "Shot noise in a strange metal"}
\author{V. R. Shaginyan}\email{vrshag@thd.pnpi.spb.ru} \affiliation{Petersburg
Nuclear Physics Institute,  Gatchina, 188300,
Russia}\affiliation{Clark Atlanta University, Atlanta, GA 30314,
USA}\author{A. Z. Msezane}\affiliation{Clark Atlanta University,
Atlanta, GA 30314, USA}\author{G. S. Japaridze}\affiliation{Clark
Atlanta University, Atlanta, GA 30314, USA} \maketitle

The recent paper \cite{science} is devoted to measurements of shot
noise to probe excitations in nanowires of the heavy fermion (HF)
metal $\rm YbRh_2Si_2$. The authors observed that shot noise is
strongly suppressed, and claim that the suppression cannot be
attributed to either electron-phonon or electron-electron
interactions in a Fermi liquid. Their observation suggests that the
current is not carried by well-defined quasiparticles in the $\rm
YbRh_2Si_2$, and calls for similar research into other strange
metals.

In this comment, we show that it is unlikely that the affected
carriers in bulk $\rm YbRh_2Si_2$ would have undergone any
fragmentation. Rather, it is the electron-phonon interaction that
suppresses noise. Indeed, experimental observations unambiguously
show that the Wiedemann-Franz law holds in $\rm YbRh_2Si_2$,
implying that no fundamental breakdown of quasiparticle behavior
occurs in the archetypical HF metal \cite{petr}. Moreover, detailed
study of the temperature evolution of quasiparticles in the HF
metal $\rm Sr_2RuO_4$ demonstrates that quasiparticles persist up
to temperatures above 200 K, far beyond the Fermi liquid regime
\cite{prl}.

It was shown  that HF metals can exhibit a quasiclassical behavior
that remains applicable to the description of the resistivity of HF
metals due to the presence of a transverse zero-sound collective
mode, reminiscent of the phonon mode in solids. It is demonstrated
that at temperatures $T$,  in excess of an extremely low Debye
temperature $T_D\lesssim 1$ K, the resistivity $\rho(T)\propto T$,
since the mechanism, forming this dependence, is the same as the
electron-phonon mechanism that prevails at high temperatures in
ordinary metals \cite{srruo,atom}. The same mechanism allows one to
explain the optical conductivity in $\rm YbRh_2Si_2$ \cite{mdpi},
and suppressing the short noise in HF metals. It is worth noting
that the current-carrying excitations in nanowires of  $\rm
YbRh_2Si_2$ can be different from the excitation of the bulk $\rm
YbRh_2Si_2$, since the nanowires can be represented by one
dimensional HF metal that possesses special properties, see e.g.
\cite{physrep}.

We conclude that there is no reason to assume the fragmentation of
charge carriers, and we can refer to William Shakespeare's immortal
comedy that  the paper \cite{science} is ``Much Ado About
Nothing.''

\end{document}